\documentclass[10pt,twocolumn,english]{paper}
\usepackage[T1]{fontenc}
\usepackage[latin9]{inputenc}
\usepackage[letterpaper]{geometry}
\usepackage{color}
\usepackage{babel}

\usepackage{amsmath}
\usepackage{graphicx}
\usepackage{amssymb}
\usepackage[unicode=true, 
 bookmarks=true,bookmarksnumbered=true,bookmarksopen=true,bookmarksopenlevel=1,
 breaklinks=true,pdfborder={0 0 0},backref=page,colorlinks=true]
 {hyperref}
\hypersetup{
 pdfauthor={Sergio Tafur},
 linkcolor=midnightblue, citecolor=midnightblue}

\makeatletter

\newcommand{\lyxdot}{.}

\usepackage{times}

\usepackage{color}
\definecolor{orange}{rgb}{1,0.5,0}
\definecolor{midnightblue}{rgb}{0,0.1,0.3}

\newcommand{\Feyn}[1]{#1\kern-0.45em/}

\usepackage{calligra}
\DeclareMathAlphabet{\squigle}{T1}{calligra}{m}{n}
\DeclareFontShape{T1}{calligra}{m}{n}{<->s*[2.2]callig15}{}

\@ifundefined{showcaptionsetup}{}{%
 \PassOptionsToPackage{caption=false}{subfig}}
\usepackage{subfig}
\makeatother

\begin{document}

\title{Single Photon Near Field Emission and Revival in Quantum Dots}

\author{Sergio Tafur%
\thanks{Nanoscience Technology Center, Department of Physics, and Institute
for Simulation and Training at the University of Central Florida,
e-mail: stafur@mail.ucf.edu%
} ~and Michael N. Leuenberger%
\thanks{ Nanoscience Technology Center and Department of Physics at the University
of Central Florida, e-mail: mleuenbe@mail.ucf.edu, corresponding author%
}}

\maketitle
\noindent \textbf{\footnotesize Models of the spontaneous emission
of photons coupled to the electronic states of quantum dots are important
for understanding quantum interactions in dielectric media as applied
to proposed solid-state quantum computers, single photon emitters,
and single photon detectors.  The characteristic lifetime of photon
emission is traditionally modeled in the Weisskopf-Wigner approximation.
Here we model the fully quantized spontaneous emission, including
near field effects, of a photon from the excited state of a quantum
dot beyond the Weisskopf-Wigner approximation. We propose the use
of discretized central-difference approximations  to describe single
photon states via single photon operators in 3+1 dimensions.  We
further show herein that one can shift from the traditional description
of electrodynamics and quantum electrodynamics, in terms of electric
and magnetic fields, to one in terms of a photonic wave function
and its operators using the Dirac equation for the propagation of
single photons.}{\footnotesize \par}

The field of quantum computation (QC) and quantum information technology
(QIT) has recently experienced escalated activity in the search for
sources of photonic states coupled to their quantum sources \cite{Imamoglu-15-Nov-1999,seigneur2008single}.
The observed increase in activity is in part due to the suggestion
that quantum information processing based on the electron spins of
quantum dots (QDs), coupled through optical modes of a micro-cavity,
\cite{Imamoglu-15-Nov-1999} could improve on schemes based on the
energy states of trapped ions \cite{cirac1995quantum} and nuclear
spins in chemical solution \cite{cory1997ensemble,gershenfeld1997bulk}.
Some advantages of a QC scheme based on the suggested semiconductor
quantum dot arrays may include greater scalability, longer spin decoherence
times, longer coherence lengths, and fast interactions mediated by
photons \cite{Imamoglu-15-Nov-1999}. In this contribution we present
a model for describing the coupling between a single photonic state
to the spin states of a quantum source (such as a QD) through optical
modes present in a micro-cavity. Additional applications of this model
may include the design of devices aimed at single photon emission
\cite{yuan2002electrically}, single photon detection \cite{komiyama2000single,astafiev2002single},
quantum teleportation \cite{bouwmeester1997experimental,bennett1993teleporting},
quantum computing within a quantum network \cite{seigneur2008single},
and quantum cryptography \cite{bennett1984quantum,ekert1991quantum,naik2000entangled,ursin2007entanglement}.
This model requires a description of optical modes present in photonic
crystals and dielectric micro-cavities. For example, in order to successfully
describe the entanglement between photons and their quantum sources,
it is imperative to achieve a resolution high enough to describe whispering
gallery modes \cite{mccall1992whispering} available in dielectric
micro-cavities such as micro-disks. This is especially important
in applications that contrast classical computers, which depend on
bits to store and process information, to quantum computers that depend
on \underbar{qu}antum \underbar{bits} (qubits). 

As the name suggests, quantum binary digits are a quantum representation
of the on and off state as interpreted in machines like ENIAC%
\footnote{Electronic Numerical Integrator And Computer%
} and those in existence today. It is possible to visualize the relationship
between bits and qubits by means of what is known as the Bloch sphere
 in terms of the on $\left|1\right\rangle $ and off states $\left|0\right\rangle $
of a bit and the state of a qubit $\left|\Psi\right\rangle $. One
major advantage of quantum computers is that they do not alter the
Church-Turing thesis \cite{nielsen2002quantum} since the do not allow
the computation of functions which are not theoretically computable.
So far it has been shown through the discoveries of Shor and others
that it is possible to develop quantum algorithms for important problems
like prime factorization \cite{shor1994algorithms}, protocols for
quantum error correction (QEC), and fault-tolerant QC \cite{shor1996fault}.
Other algorithms in QC, that if physically implemented could be of
immediate use, include Grover's Algorithm for database searches \cite{grover1996fast}
and the quantum Fourier Transform \cite{hales2000improved}. Also
central to the discussion on QC and QEC is the decoherence rate of
qubits. It is imperative to QC to find an implementation where qubits
are well isolated from their environment \cite{nielsen2002quantum}.
Among suggested implementations for QC are Raman coupled low-energy
states of trapped ions \cite{cirac1995quantum} and nuclear spins
in chemical solution \cite{cory1997ensemble,gershenfeld1997bulk}.
Qubits based on these implementations could provide the first examples
of QC up to the 5 through 10 qubits level. However, these implementations
may not be scalable to more than 100 qubits \cite{Imamoglu-15-Nov-1999}. 

 Proposed implementations and promising schemes that could be scalable
to more than 100 coupled qubits may be based on electron spins coupled
by means of an optical mode of a photonic crystal or dielectric micro-cavity.
One such scheme couples the electronic spin states of a Quantum Dot
(QD) to the optical modes of a micro-disk \cite{Imamoglu-15-Nov-1999}.
Another promising scheme couples electronic QD states embedded inside
nanocavities to the modes of a photonic crystal host \cite{leuenberger2005teleportation,leuenberger2006fault,gonzalez2010theory}.
An additional scheme recently realized experimentally has shown that
Nitrogen-Vacancy centers in diamond can also be embedded inside a
photonic crystal, which could enable fully scalable room-temperature
quantum computing as a good alternative to using quantum dots. \cite{awschalom2007diamond,wang2007fabrication,gonzalez2010dynamics,gonzalez2009theory}. 

\begin{figure}
\begin{minipage}[t]{1.7in}%
\subfloat[Heavy-Heavy hole]{\centering{}\includegraphics[width=1\columnwidth]{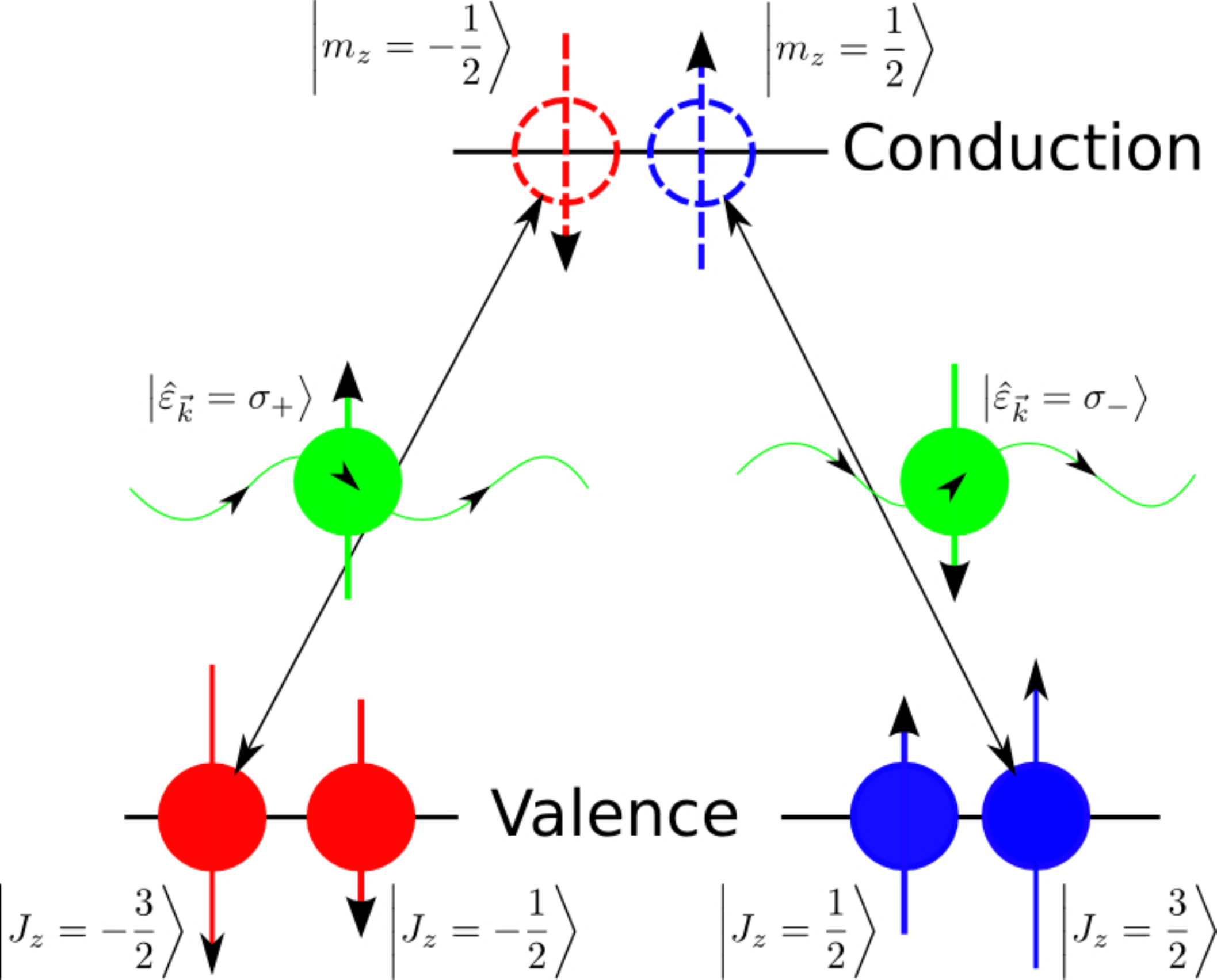}}%
\end{minipage}\hfill{}%
\begin{minipage}[t]{1.7in}%
\subfloat[Light-Light hole]{\centering{}\includegraphics[width=1\columnwidth]{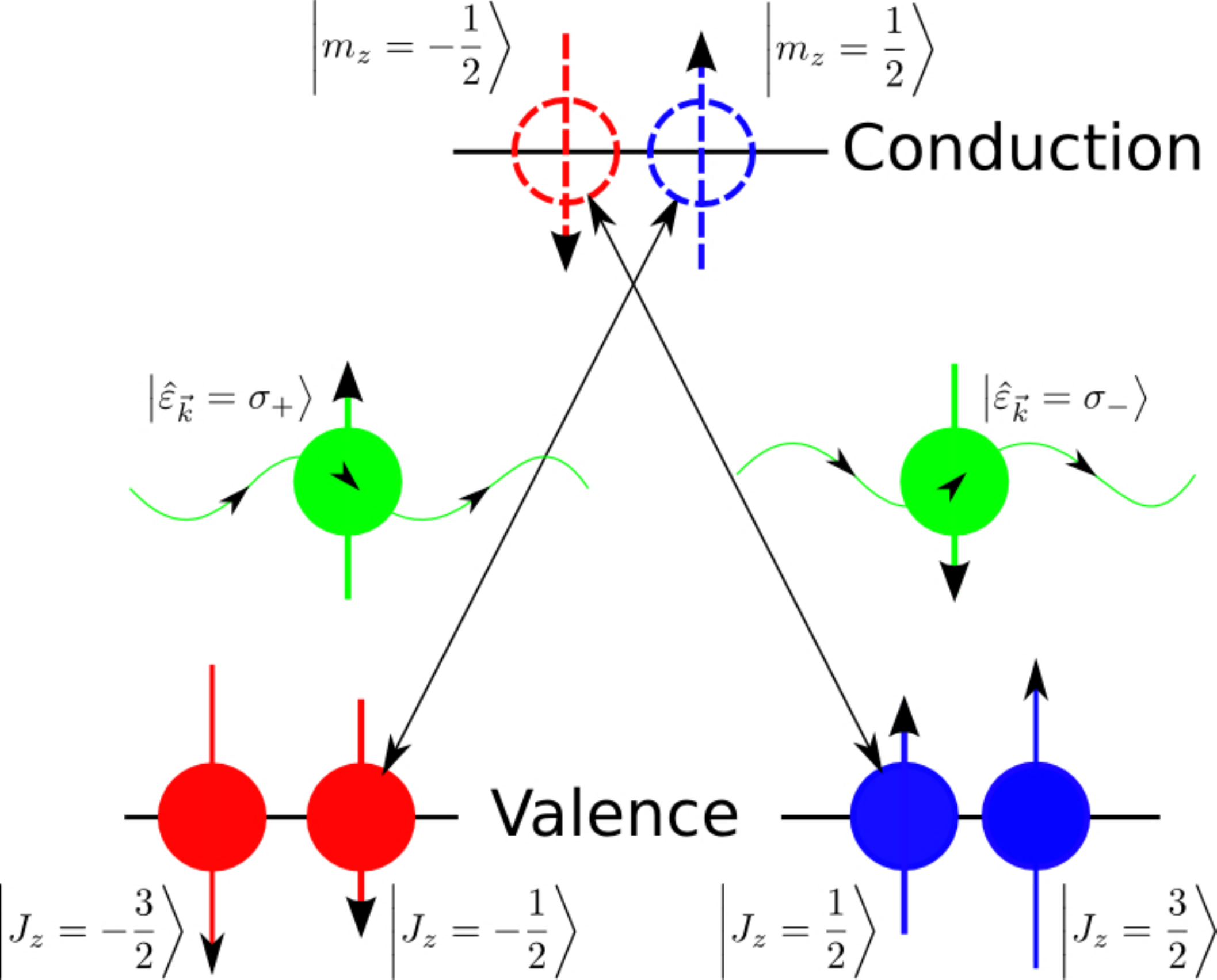}}%
\end{minipage}\caption{Selection rules in a two-level quantum source coupling to single photon
states\label{fig:Selection-rules}}

\end{figure}

The discussion for generating a theory on the interaction between
such quantum sources and photonic states may be modeled after atomic
systems \cite{Quantum-Optics-Scully}. For the case of a quantum source
modeled after an effective two-level QD, the selection rules presented
in figure \eqref{fig:Selection-rules} yield an interaction via the
dipole approximation by implementing the state to state transition
dipole moments of the QD \cite{seigneur2008single}. To this end it
is imperative to model spontaneous emission of a photon coupled to
the electronic state of a quantum dot beyond the Weisskopf-Wigner
approximation; which fixes the value of accessible modes from a possibly
infinite set of frequencies $\nu_{k}$ to a single mode of a cavity
or the transition frequency of the QD state as represented by $\omega_{\gamma}$
\cite{Quantum-Optics-Scully}.%

 Our proposed model describes the photon by means of a Dirac-like
equation for the photon. Experimentally, the quantum state of a photon
may be reconstructed using optical homodyne tomography techniques
by measuring quantum noise statistics of field amplitudes at different
optical phases \cite{leonhardt1995quantum,lvovsky2009continuous}.
In this work, the rigorous description of the interaction between
a quantum source and the generated Maxwell Field is initially made
within the formalism of relativistic quantum field theory (QFT). In
this description we begin within the canonical quantization procedure
presented by the Gupta and Beuler method  and the resulting interaction
between these fields \cite{bjorken-rqf1965}. This procedure requires
the definition of a Lagrangian and gauge for the interacting fields%
\footnote{In the relativistic regime we set ${\displaystyle \vec{\nabla}\cdot\vec{A}+\frac{1}{c}\partial_{t}\Phi=0}$.%
}. To follow this procedure, a connection between the photon wave function
(PWF) \cite{PhysRev.105.1914,moses1959solution,bialynicki1994wave,smith2007photon}
 and the four vector potential for a Maxwell Field has to be drawn.
This canonical quantization procedure leads to two important results,
the complex Maxwell Field Tensor and the coupled electron-photon field
equations in terms of a field equation for the PWF. We show the first
result to be a Lagrangian for a free complex Maxwell Field written
in terms of a self-dual tensor representing the field tensor corresponding
to the PWF that directly satisfies and yields equations of motion
equivalent to the generalized Maxwell equations \cite{gersten2001maxwell,Varlamov2005MaxwellField,bialynicki1996v,smith2007photon}.
In the second result we show the coupling between a quantum source
and the Maxwell Field it generates. We additionally extend these results
to show how these can be applied to model the emission of a single
photon from a dielectric micro-cavity. Throughout the relativistic
treatment of these fields we will maintain the Minkowski Metric to
have the signature $\left(+,-,-,-\right)$, and adopt the 4-notation
consistent with $x^{\mu}\equiv\left(ct,\vec{x}\right)$, $\eta_{\mu\nu}x^{\nu}=x_{\mu}=\left(ct,-\vec{x}\right)$.

Building on the work describing the PWF formalism, while working in
Gaussian units in the presence of sources, one can define a self-dual
tensor in terms of the electromagnetic field tensor \cite{gersten2001maxwell}
$\mathfrak{F}^{\mu\nu}=\partial^{\mu}A^{\nu}-\partial^{\nu}A^{\mu}$
 and its dual $\mathfrak{F}_{\mu\nu}^{D}=\frac{1}{2}\varepsilon_{\mu\nu\alpha\beta}\mathfrak{F}^{\alpha\beta}$
as $\mathfrak{G}^{\mu\nu}\equiv\mathfrak{F}^{\mu\nu}-i\mathfrak{F}_{D}^{\mu\nu}$
in terms of the vector potential by \begin{flalign*}
 & \mathfrak{G}^{\mu\nu}=\left(\partial^{\mu}A^{\nu}-\partial^{\nu}A^{\mu}\right)-\frac{i}{2}\varepsilon^{\mu\nu\alpha\beta}\left(\partial_{\mu}A_{\nu}-\partial_{\nu}A_{\mu}\right)\end{flalign*}
from which it is possible to define a general gauge-invariant Lagrangian
for a free photon via\begin{flalign*}
 & \mathfrak{L}_{\textrm{photon}}=-\frac{1}{8}\mathfrak{G}_{\mu\nu}\mathfrak{G}^{\mu\nu}\end{flalign*}
The same Lagrangian expressed in terms of the well known Faraday tensor\cite{landau1980classical}
reads\begin{flalign*}
 & \mathfrak{L}_{\textrm{photon}}=-\frac{1}{8}\left(\mathfrak{F}_{\mu\nu}\mathfrak{F}^{\mu\nu}+i\left(\mathfrak{F}_{\mu\nu}\mathfrak{F}_{D}^{\mu\nu}+\mathfrak{F}_{\mu\nu}^{D}\mathfrak{F}^{\mu\nu}\right)+\mathfrak{F}_{\mu\nu}^{D}\mathfrak{F}_{D}^{\mu\nu}\right)\end{flalign*}
with the corresponding matrix representation for the self-dual tensor
given by \begin{flalign*}
 & \mathfrak{G}^{\mu\nu}=\left(\begin{array}{cccc}
0 & i\digamma_{x}^{+} & i\digamma_{y}^{+} & i\digamma_{z}^{+}\\
-i\digamma_{x}^{+} & 0 & -\digamma_{z}^{+} & \digamma_{y}^{+}\\
-i\digamma_{y}^{+} & \digamma_{z}^{+} & 0 & -\digamma_{x}^{+}\\
-i\digamma_{z}^{+} & -\digamma_{y}^{+} & \digamma_{z}^{+} & 0\end{array}\right)\end{flalign*}
Using $\vec{E}$ and $\vec{B}$ to represent electric and magnetic
fields respectively \cite{griffiths1999introduction} , {\normalsize $\vec{\digamma}_{\mp}\equiv\vec{B}\pm i\vec{E}$
}\cite{moses1959solution,PhysRev.105.1914,landau1980classical}. This
vector may be written in terms of the Riemann-Silberstein vector \cite{bialynicki1994wave,smith2007photon}
through the identity $\vec{\mathcal{F}}_{\pm}\equiv\pm i\vec{\digamma}_{\pm}=\vec{E}\pm i\vec{B}$
(given that $i$ represents the unit pseudo-scalar \cite{de2007geometric}).

We defer to the traditional Quantum Electro-Dynamics (QED) interaction
between an electron, a photon, and an additional gauge field as mitigated
by $\bar{\psi}\Feyn{A}_{\text{Tot}}\psi$ where $\left(\Feyn{\Lambda}\equiv\gamma^{\mu}\Lambda_{\mu}\right)$,
via $\Feyn{A}_{\text{Tot}}\equiv\Feyn{A}+\Feyn{A}_{\text{Ext}}$ and
therefore express the interaction Lagrangian for a gauge field $A_{\mu}$
and spinor $\psi\equiv\left(\begin{array}{c}
\varphi\\
\chi\end{array}\right)$ similar to the the description in \cite{huang2010quantum} by writing 

\begin{flalign*}
 & \mathcal{L}_{\text{Int}}=\bar{\psi}\left(\Feyn{p}-m_{0}c\right)\psi-\frac{e}{c}\bar{\psi}\left(\Feyn{A}_{\text{Tot}}\right)\psi-\frac{1}{8}\mathfrak{G}_{\mu\nu}\mathfrak{G}^{\mu\nu}\end{flalign*}
It is worth noting that this Lagrangian leads to electromagnetic fields
that satisfy the principle of superposition as required by experiment
along with their conservation laws and definition of spin. This is
evident because there are only expressions quadratic in the field
and first order time derivatives present in the action. Explicitly
in terms of the relativistic equations of motion and their Hermitian
conjugates \begin{flalign}
 & \left[i\hbar\gamma^{\mu}\partial_{\mu}+\frac{e}{c}\gamma^{\mu}A_{\mu,\text{Tot}}+m_{0}c\right]\bar{\psi}=0\\
 & -\frac{e}{c}\bar{\psi}\gamma^{\nu}\psi+\partial_{\mu}\mathfrak{G}^{\mu\nu}=0\end{flalign}
In the non-relativistic limit working in the radiation gauge, these
EOMs yield the well known Pauli-Schr\"{o}dinger equation \cite{sakurai2006advanced}.
By quantizing the 3-vector potential while defining the new operators
$\vec{\digamma}_{+}=\vec{\nabla}\times\vec{A}-i\frac{1}{c}\partial_{t}\vec{A}$
\& $\vec{\digamma}_{-}=\vec{\nabla}\times\vec{A}+i\frac{1}{c}\partial_{t}\vec{A}$,
the EOM for the photon may be expressed as a Dirac-like equation \begin{flalign}
 & \left[\frac{i\hbar}{c}\partial_{t}\left(\begin{array}{cc}
0 & I\\
-I & 0\end{array}\right)-\frac{\hbar}{i}\partial_{k}\left(\begin{array}{cc}
0 & \sigma_{k}^{\left(3\right)}\\
\sigma_{k}^{\left(3\right)} & 0\end{array}\right)\right]\left(\begin{array}{c}
\vec{\digamma}_{+}\\
\vec{\digamma}_{-}\end{array}\right)=0\label{eq:Dirac-like-photon-eqn}\end{flalign}
where $\sigma_{k}^{\left(3\right)}=-i\varepsilon_{ijk}$ with $\varepsilon_{ijk}$
representing the Levi-Civita permutation symbol. To define the quantized
interaction term we expand the gauge field operator $\vec{A}$ in
terms of creation and annihilation operators through the use of plane
waves and retain a phase factor $\phi$ to account for its phase freedom
\cite{smith2007photon}

\begin{flalign*}
 & \vec{A}=\sum_{\vec{k},\lambda}\frac{c}{\nu_{k}}\sqrt{\frac{\hbar\nu_{k}}{2V}}\left(\hat{\epsilon}_{\vec{k},\lambda}a_{\vec{k},\lambda}e^{-i\left(\nu_{k}t-\vec{k}\cdot\vec{x}\right)}e^{-i\phi}+\text{H.c.}\right)\end{flalign*}
Making the dipole approximation to the Pauli-Schr\"{o}dinger equation
\cite{sakurai2006advanced}, and changing to the interaction picture
leads to the expression \begin{flalign*}
 & i\hbar\partial_{t}\left|\varphi\right\rangle =e^{i\frac{\pi}{2}}\sum_{n,m,\vec{k},\lambda}\sqrt{\frac{\hbar\nu_{k}}{2V}}\left(\hat{\epsilon}_{\vec{k},\lambda}a_{\vec{k},\lambda}e^{-i\left(\nu_{k}t-i\vec{k}\cdot\vec{x}_{0}\right)}e^{-i\phi}\right.\\
 & +\left.\hat{\epsilon}_{\vec{k},\lambda}^{*}a_{\vec{k},\lambda}^{\dagger}e^{i\left(\nu_{k}t-i\vec{k}\cdot\vec{x}_{0}\right)}e^{i\phi}\right)\cdot\left(\vec{\wp}_{nm}\sigma_{nm}e^{i\omega_{nm}t}\right)\left|\varphi\right\rangle \end{flalign*}
Assuming that $\nu_{k}=c\left|k\right|$ and making use of the identity
\cite{Quantum-Optics-Scully} ${\displaystyle \frac{\vec{k}}{k}\times\hat{\epsilon}_{\vec{k},\lambda}}=-\sigma i\hat{\epsilon}_{\vec{k},\lambda}$,
yields that the expression for the interaction can be written in terms
of the operator $\vec{\digamma}_{+}$ \begin{flalign*}
 & \vec{\digamma}_{+}=e^{-i\frac{\pi}{2}}\sum_{\vec{k}}\sqrt{\frac{2\hbar\nu_{k}}{V}}\left(\hat{\epsilon}_{\vec{k},+}a_{\vec{k},+}e^{-i\left(\nu_{k}t-\vec{k}\cdot\vec{x}\right)}e^{-i\left(\phi-\frac{\pi}{2}\right)}\right.\\
 & +\left.\hat{\epsilon}_{\vec{k},-}^{*}a_{\vec{k},-}^{\dagger}e^{i\left(\nu_{k}t-\vec{k}\cdot\vec{x}\right)}e^{i\left(\phi-\frac{\pi}{2}\right)}\right)\end{flalign*}
by shifting the phase of the interaction $\phi\rightarrow\phi-\frac{\pi}{2}$,
such that at $\vec{x}_{0}$, \begin{flalign}
 & i\hbar\partial_{t}\left|\varphi\right\rangle =-\frac{1}{2}\sum_{n,m}\left(\vec{\digamma}_{+}-\vec{\digamma}_{-}\right)\cdot\left(\vec{\wp}_{nm}\sigma_{nm}e^{i\omega_{nm}t}\right)\left|\varphi\right\rangle \label{eq:GeneralInteractionPicture-digamma}\end{flalign}
The EOMs as derived from \eqref{eq:GeneralInteractionPicture-digamma}
(in terms of the photonic wave functions), for the case of a two level
quantum source with an energy band-gap of $\Delta E=\hbar\omega_{\sigma}$,
defined by the state-vector%
\footnote{Where $\sigma$, $\gamma$ denote electronic and photonic states and
$a$, $b$ denote excited and ground states%
} $\left|\sigma\gamma\right\rangle =c_{a}\left(t\right)\left|a0\right\rangle +c_{b,\vec{k}}\left(t\right)\left|b1_{\vec{k}}\right\rangle $
interacting with it's own spontaneously emitted field, are given by%
\footnote{e.g. $\vec{\Psi}_{\gamma,+,b}^{\left(+\right)}=e^{i\frac{\pi}{2}}{\displaystyle \sum_{\vec{k}}}\sqrt{\frac{\hbar\nu_{k}}{2V}}c_{b,\vec{k},+}\left(t\right)\hat{\epsilon}_{\vec{k},+}e^{-i\nu_{k}t}e^{i\vec{k}\cdot\vec{x}_{0}}e^{-i\left(\phi-\frac{\pi}{2}\right)}$
\& $\vec{\Psi}_{\gamma,-,b}^{*\left(+\right)}=e^{i\frac{\pi}{2}}{\displaystyle \sum_{\vec{k}}}\sqrt{\frac{\hbar\nu_{k}}{2V}}c_{b,\vec{k},-}\left(t\right)\hat{\epsilon}_{\vec{k},-}e^{-i\nu_{k}t}e^{i\vec{k}\cdot\vec{x}_{0}}e^{-i\left(\phi-\frac{\pi}{2}\right)}$%
}\begin{flalign*}
 & i\hbar\dot{c}_{a}\left(t\right)=\left(\vec{\Psi}_{\gamma,+}^{\left(+\right)}+\vec{\Psi}_{\gamma,-}^{*\left(+\right)}\right)_{,b}e^{i\omega_{\sigma}t}\cdot\vec{\wp}_{ba}\\
 & i\hbar\sum_{\vec{k},\pm}\dot{c}_{b,\vec{k},\pm}\left(t\right)=\left(\vec{\Psi}_{\gamma,+}^{\left(-\right)}+\vec{\Psi}_{\gamma,-}^{*\left(-\right)}\right)_{,a}e^{-i\omega_{\sigma}t}\cdot\vec{\wp}_{ab}\end{flalign*}
Formally integrating these equations of motion while representing
the polarization vectors as $\hat{\epsilon}_{\vec{k},+}=\frac{1}{\sqrt{2}}\left(\hat{\theta}+i\hat{\phi}\right)$,
$\hat{\epsilon}_{\vec{k},-}=\frac{1}{\sqrt{2}}\left(\hat{\theta}-i\hat{\phi}\right)$,
and the unit wave-vector as $\hat{k}\equiv\hat{r}$, yield three poles
$z_{n}$ which need to be considered when solving for the probability
density that ultimately gives rise to the spontaneously emitted photon,
$c_{b,\vec{k},\lambda}\left(t\right)$. Each of these poles correspond
to emission and re-absorption respectively and allow one to model
Rabi oscillations associated with revival phenomena. These poles can
be evaluated numerically or analytically by application of Demoivre's
theorem. 

\begin{figure}
\begin{minipage}[t]{1.7in}%
\subfloat[$\theta\left(z_{0}+i\omega_{\sigma}\right)\approx-15^{\circ}$]{\centering{}\includegraphics[width=1\columnwidth]{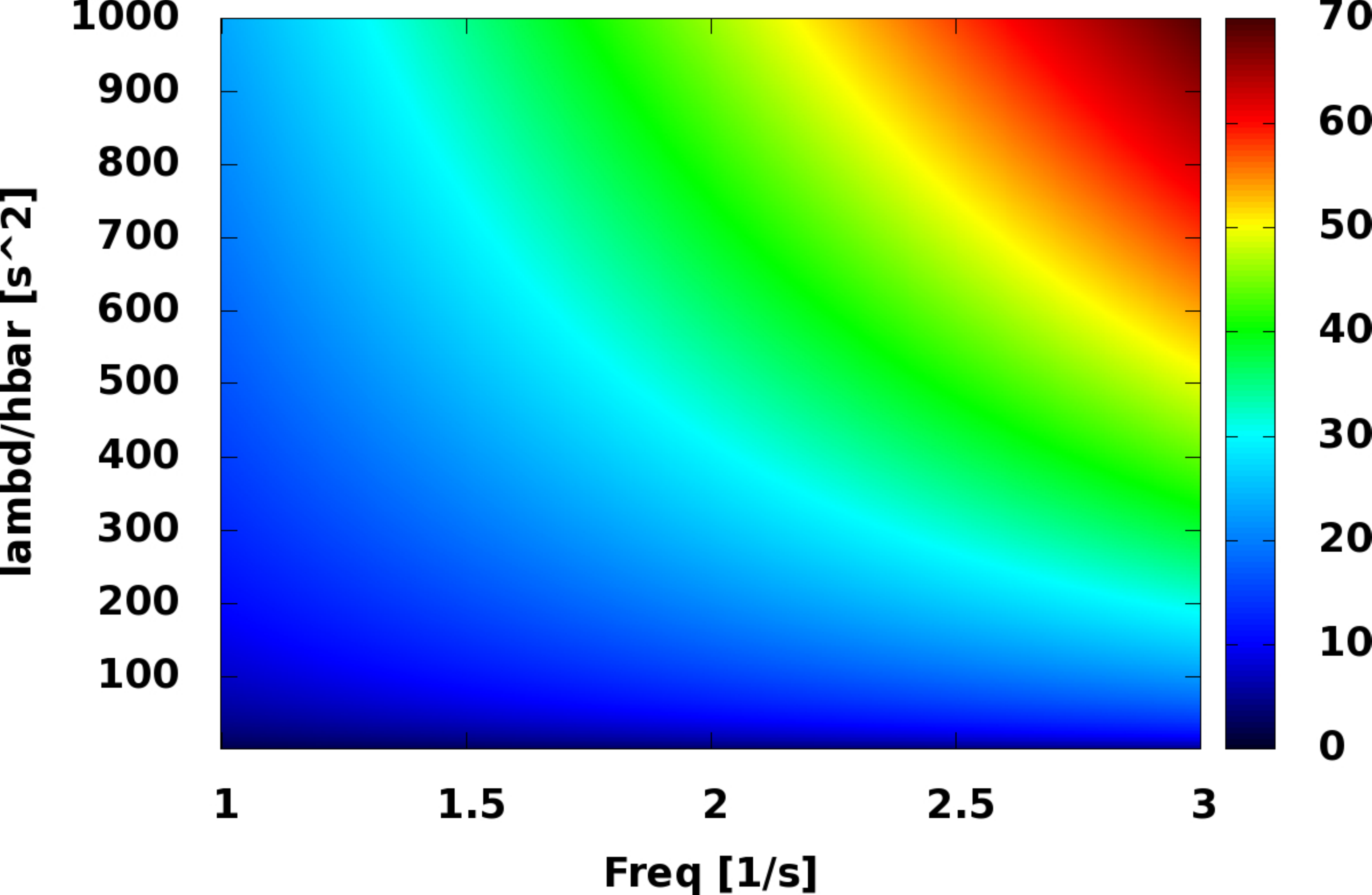}}%
\end{minipage}\hfill{}%
\begin{minipage}[t]{1.7in}%
\subfloat[$\theta\left(z_{1}+i\omega_{\sigma}\right)\approx165^{\circ}$]{\centering{}\includegraphics[width=1\columnwidth]{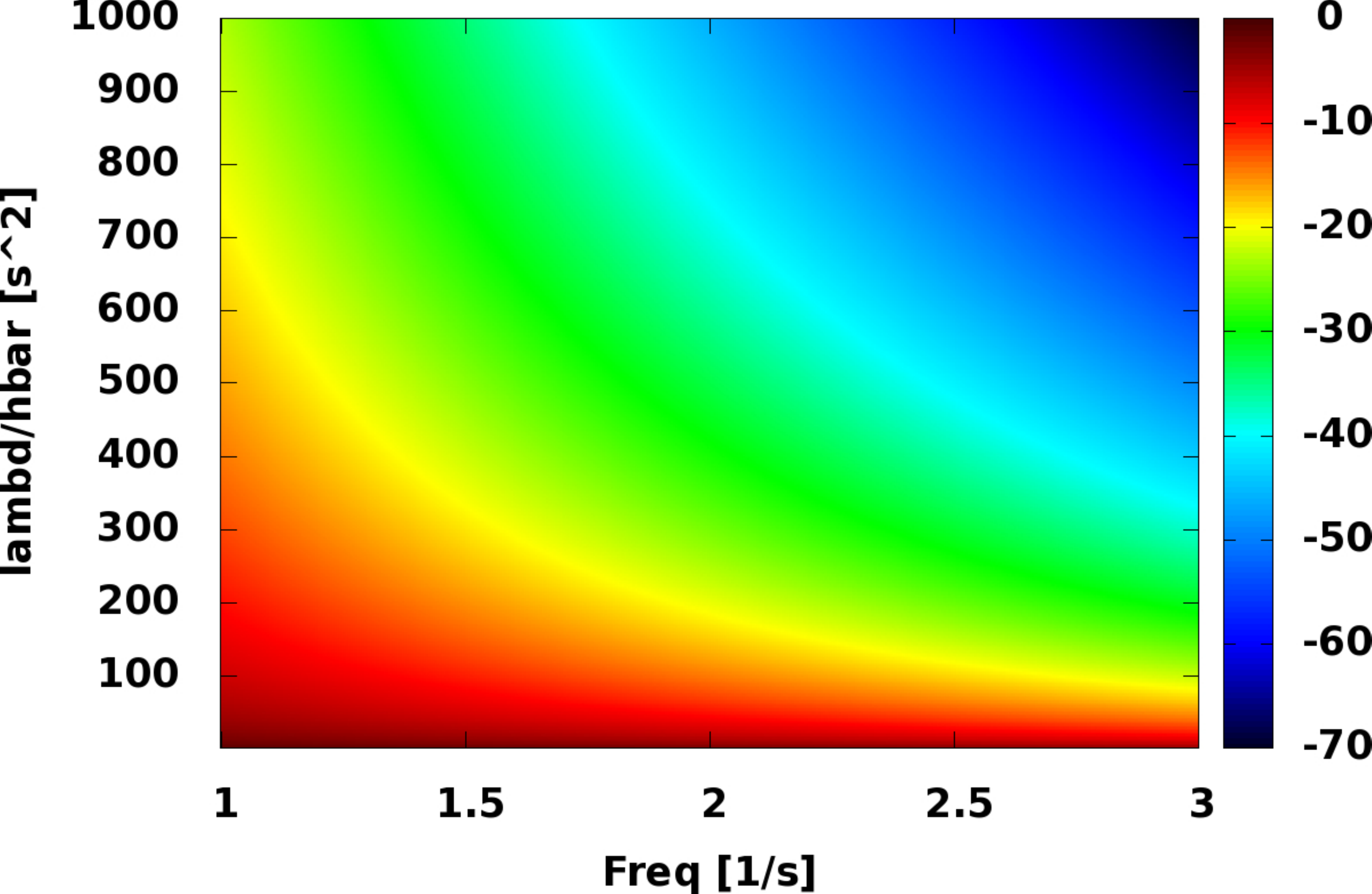}}%
\end{minipage}\hfill{}%
\begin{minipage}[t]{1.7in}%
\subfloat[$\theta\left(z_{2}+i\omega_{\sigma}\right)\approx150^{\circ}$]{\centering{}\includegraphics[width=1\columnwidth]{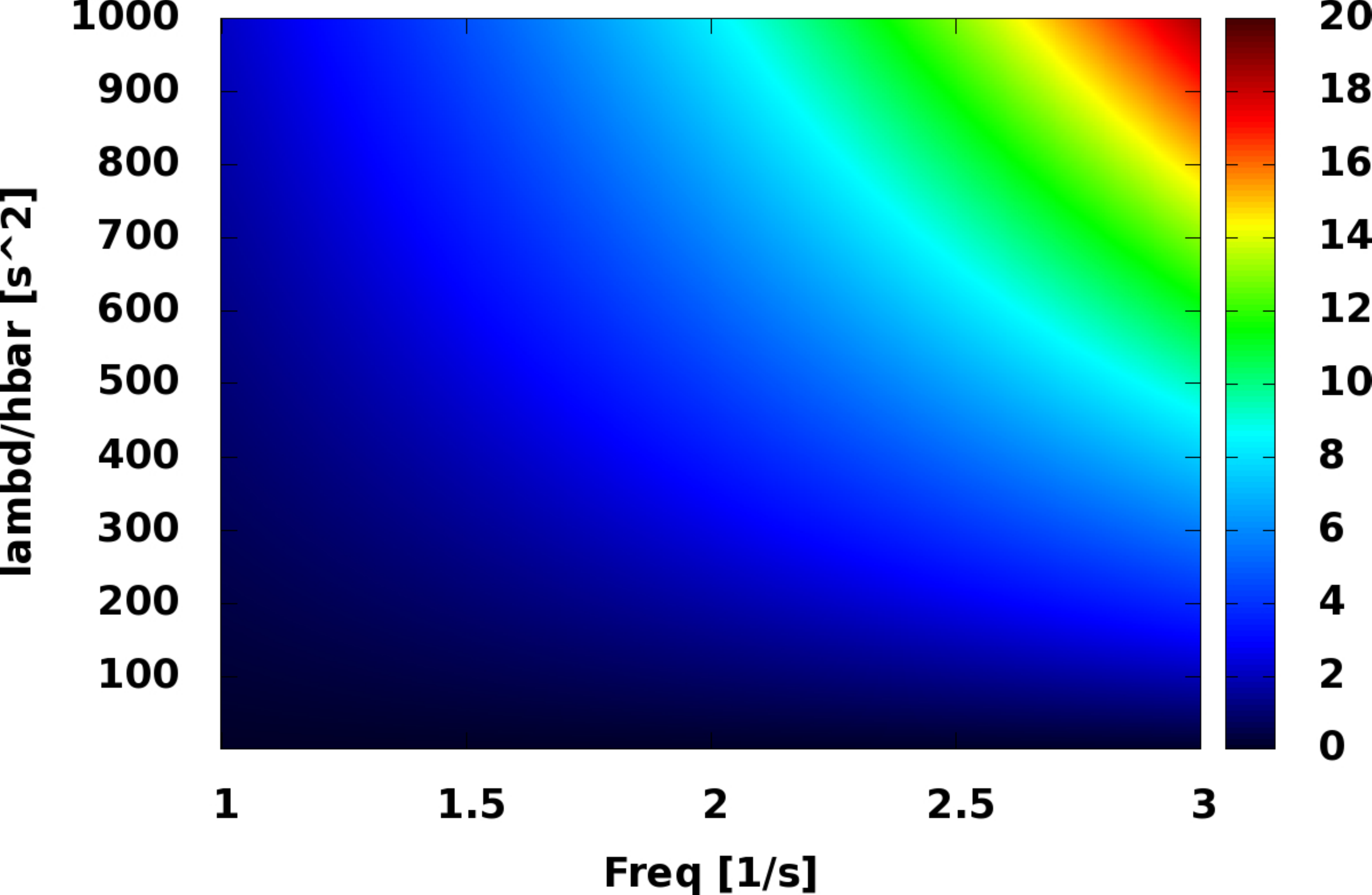}}%
\end{minipage}\hfill{}%
\begin{minipage}[t]{1.7in}%
\begin{center}
\subfloat[$\left|z_{0,1}+i\omega_{\sigma}\right|$]{\centering{}\includegraphics[width=1\columnwidth]{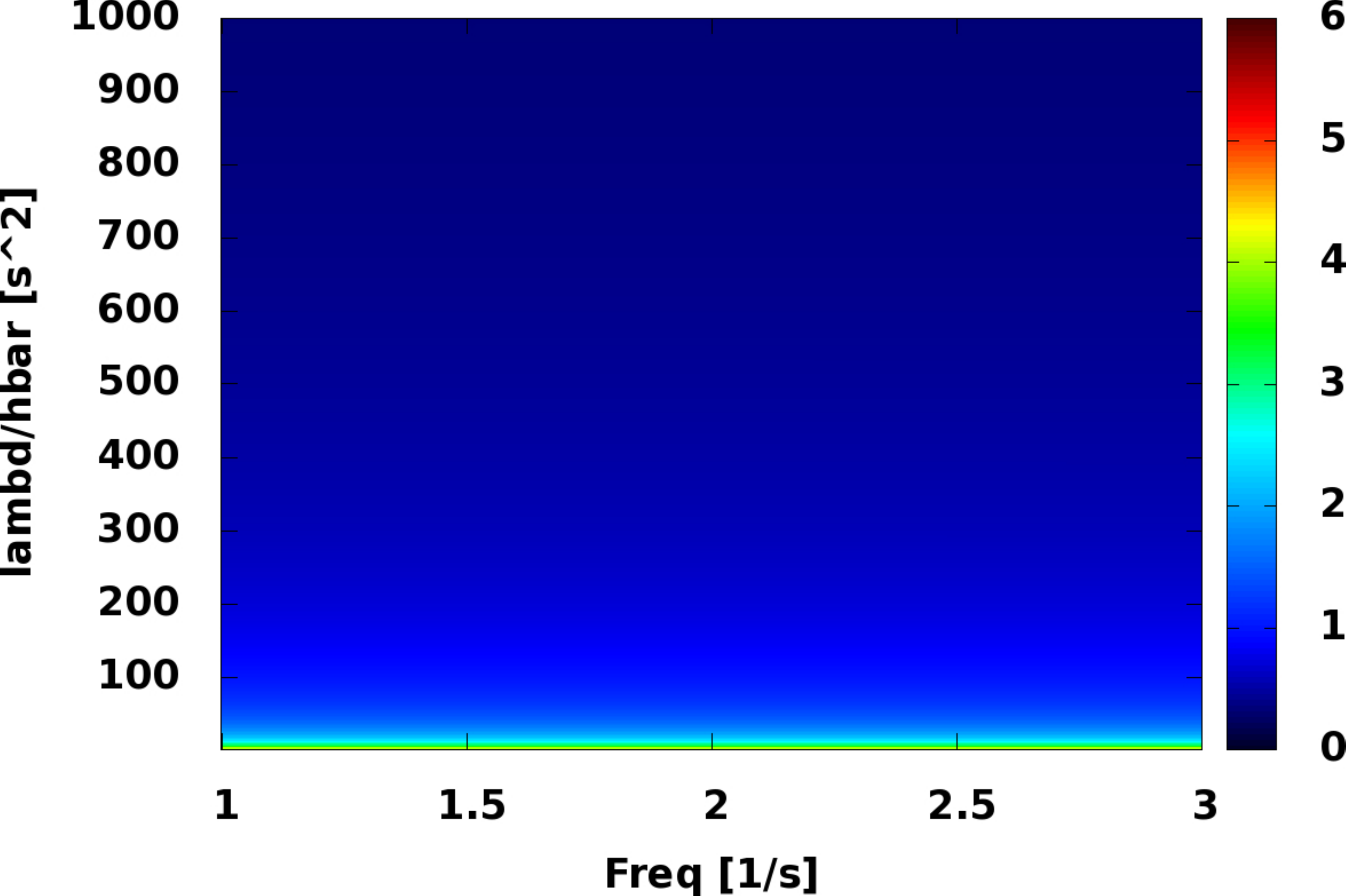}}
\par\end{center}%
\end{minipage}\hfill{}

\begin{minipage}[t]{1.7in}%
\begin{center}
\subfloat[$\left|z_{2}+i\omega_{\sigma}\right|$]{\centering{}\includegraphics[width=1\columnwidth]{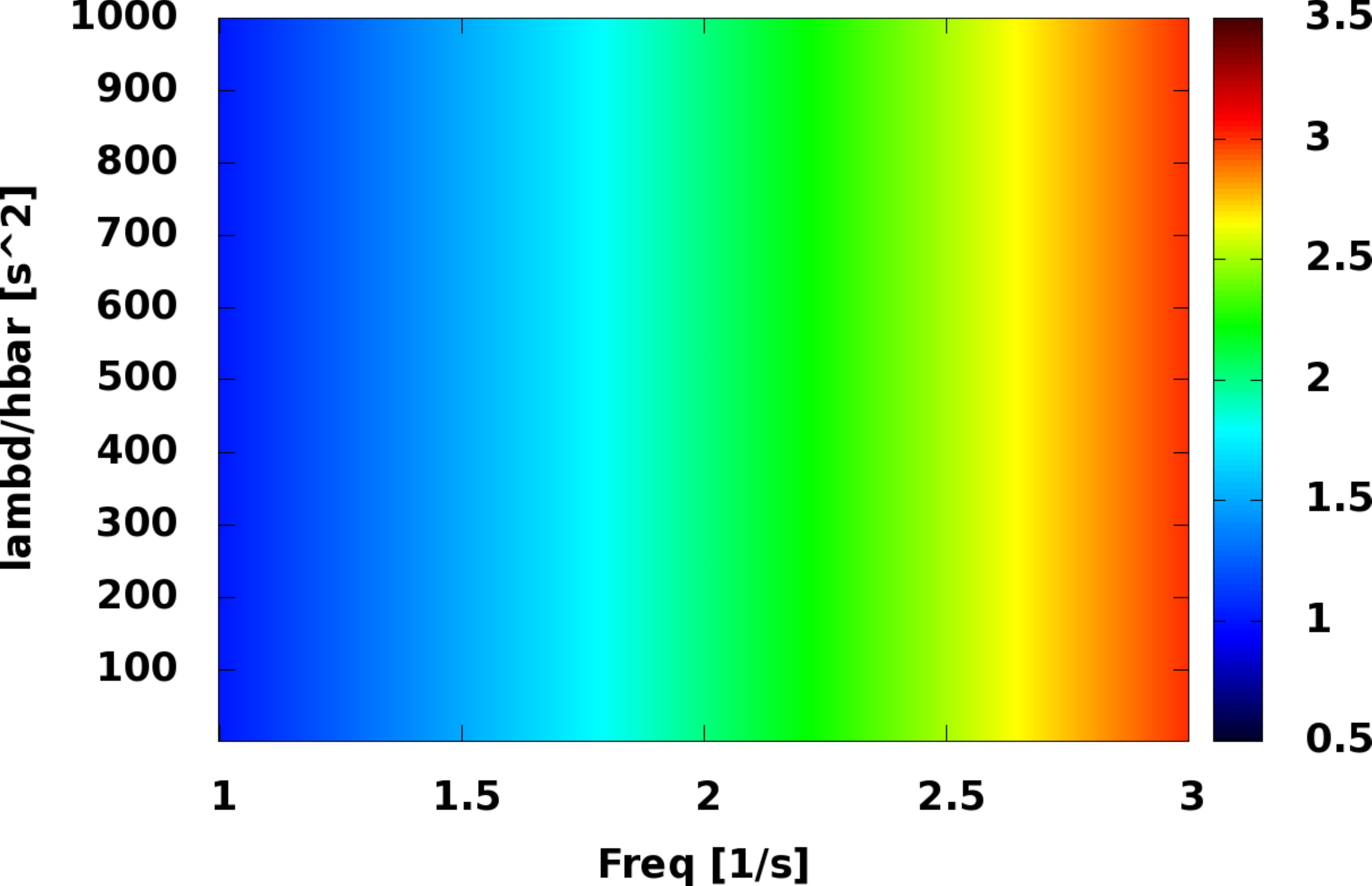}}
\par\end{center}%
\end{minipage}\hfill{}%
\begin{minipage}[t]{1.7in}%
\begin{center}
\subfloat[Phasors $z_{n}=i\omega_{\sigma}+z_{n}'$]{\centering{}\includegraphics[width=0.7\columnwidth]{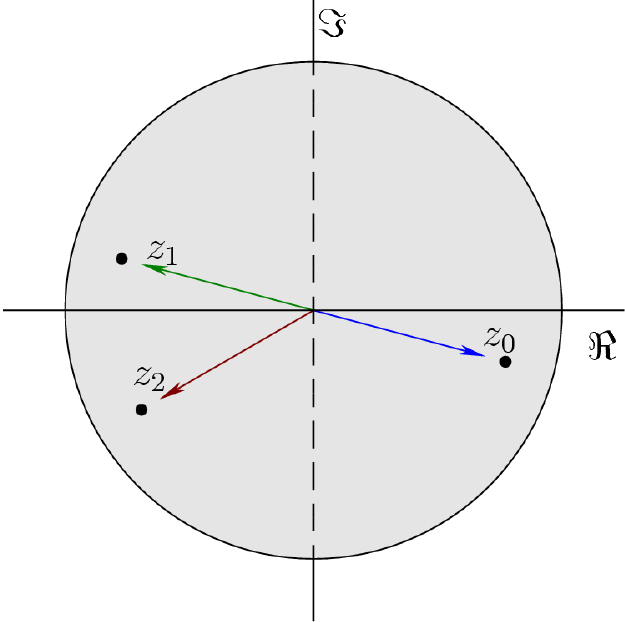}}
\par\end{center}%
\end{minipage}\caption{{\footnotesize Analytically evaluated poles $z_{n}$ through use of
Demoivre's theorem. Plots present the relative change of the argument
$\theta$ in arc-seconds $\Delta\theta^{"}$ for $z_{0}$ \& $z_{1}$.
For $z_{2}$ the change in the argument is of the order $\Delta\theta^{"}\times10^{-2}$.
Radial length phasors $z_{n}+i\omega_{\sigma}$ for $z_{0}$ and $z_{1}$
are of the order of $10^{19}\frac{1}{s}$ and for $z_{2}$ these are
of the order of $10^{15}\frac{1}{s}$ as plotted with respect to transition
frequency $\omega_{\sigma}$ in $10^{15}\frac{1}{s}$ and $\frac{\lambda}{\hbar}=\frac{1}{4\pi}\frac{1}{3\hbar c^{3}}\left[2\vec{\wp}_{ab,x}^{2}+2\vec{\wp}_{ab,y}^{2}+\vec{\wp}_{ab,z}^{2}\right]$
in $10^{-40}s^{2}$. Phasors with negative real components correspond
to emission and those with positive real components correspond to
revival in near field regions.\label{fig:Countours-over-poles}}}

\end{figure}

In figure \eqref{fig:Countours-over-poles} we present solutions for
poles $z_{n}$ in for transition wavelengths and transition dipole
moments in the ranges of 750 - 1300 nm and 20 to 100 Debye \cite{silverman2003direct,eliseev2000transition,stievater2001rabi}.
 From these results, the poles of the contour integrals associated
with the evaluation of $c_{a}\left(t\right)$ are presented as in
figure \eqref{fig:Countours-over-poles} and their physical meaning
interpreted from their location in the complex plane; where the contours
are closed with further causal constraints. The expression for the
wave-function, prior to its propagation through \eqref{eq:Dirac-like-photon-eqn}
is \begin{flalign}
 & \vec{\Psi}_{\gamma,-,b}^{*\left(+\right)}\left(t\right)=\frac{-1}{\left(2\pi\right)^{3}}\sum_{n=1}^{3}A_{n}\vec{I}_{-,n}+\vec{\Psi}_{\gamma,-,b}^{*\left(+\right)}\left(t_{0}\right)\label{eq:interaction-negative-helicity}\\
 & \vec{\Psi}_{\gamma,+,b}^{\left(+\right)}\left(t\right)=\frac{-1}{\left(2\pi\right)^{3}}\sum_{n=1}^{3}A_{n}\vec{I}_{+,n}+\vec{\Psi}_{\gamma,+,b}^{\left(+\right)}\left(t_{0}\right)\label{eq:interaction-positive-helicity}\end{flalign}
\emph{In this result we do not neglect terms of order $O\left\{ r^{-2}\right\} $},
usually neglected in the far-field approximation \cite{Quantum-Optics-Scully}.
Writing $\vec{\wp}_{ab,i}\rightarrow\vec{\wp}_{i}$ along with $\Omega_{n}=\omega_{\sigma}+iz_{n}$
and $\squigle{r}=\vec{x}-\vec{x}_{0}$, and setting the speed of light
within the interaction region to $c_{0}=\eta^{-1}c$, the components
for an {}``emitting'' pole with the contour closed over the lower
half plane, are given by %
\begin{figure}
\begin{centering}
\includegraphics[width=3.4in]{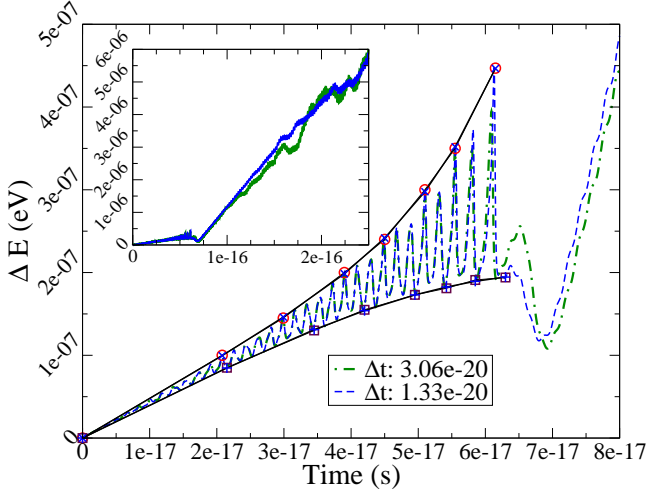}\caption{{\footnotesize Near field revival phenomena represented in terms of
energy exchange $\Delta E\,\left(eV\right)$ between the QD state
and the single photon for coarse and fine time steps. Envelope functions
bounding the region of coherent oscillation behave as low order polynomials.
The polynomial behavior of the envelope functions is contradictory
to the expectation of exponential behavior with characteristic times
of the order of the roots $z_{n}$ for $c_{a}$. \label{fig:Near-field-revival}}}

\par\end{centering}

\end{figure}
\begin{flalign*}
 & \mathcal{I}_{\pm,n,z}=\left[\left(\frac{1}{\squigle{r}^{3}}-\frac{i\Omega_{n}}{\squigle{r}^{2}c_{0}}\right)\zeta_{-}-\left(\frac{1}{\squigle{r}^{3}}+\frac{i\Omega_{n}}{\squigle{r}^{2}c_{0}}\right)\zeta_{+}\right]\vec{\wp}_{z}\end{flalign*}
\begin{flalign*}
 & \mathcal{I}_{\pm,n,x}=\\
 & \left[\left(\pm i\vec{\wp}_{y}+\frac{\left(1\pm1\right)}{2}\vec{\wp}_{x}\right)\frac{\Omega_{n}^{2}}{\squigle{r}c_{0}^{2}}+\left(\vec{\wp}_{x}\pm\vec{\wp}_{y}\right)\frac{i\Omega_{n}}{\squigle{r}^{2}c_{0}}-\frac{\vec{\wp}_{x}}{\squigle{r}^{3}}\right]\frac{\zeta_{-}}{2}\\
- & \left[\left(\pm i\vec{\wp}_{y}-\frac{\left(1\pm1\right)}{2}\vec{\wp}_{x}\right)\frac{\Omega_{n}^{2}}{\squigle{r}c_{0}^{2}}+\left(\vec{\wp}_{x}\mp\vec{\wp}_{y}\right)\frac{i\Omega_{n}}{\squigle{r}^{2}c_{0}}+\frac{\vec{\wp}_{x}}{\squigle{r}^{3}}\right]\frac{\zeta_{+}}{2}\end{flalign*}
\begin{flalign*}
 & \mathcal{I}_{\pm,n,y}=\\
 & \left[\left(\mp i\vec{\wp}_{x}-\frac{\left(1\mp1\right)}{2}\vec{\wp}_{y}\right)\frac{\Omega_{n}^{2}}{\squigle{r}c_{0}^{2}}+\left(\vec{\wp}_{y}\pm\vec{\wp}_{x}\right)\frac{i\Omega_{n}}{\squigle{r}^{2}c_{0}}-\frac{\vec{\wp}_{y}}{\squigle{r}^{3}}\right]\frac{\zeta_{-}}{2}\\
- & \left[\left(\mp i\vec{\wp}_{x}+\frac{\left(1\mp1\right)}{2}\vec{\wp}_{y}\right)\frac{\Omega_{n}^{2}}{\squigle{r}c_{0}^{2}}+\left(\vec{\wp}_{y}\mp\vec{\wp}_{x}\right)\frac{i\Omega_{n}}{\squigle{r}^{2}c_{0}}+\frac{\vec{\wp}_{y}}{\squigle{r}^{3}}\right]\frac{\zeta_{+}}{2}\end{flalign*}
as governed by the conditions that follow from the fact that the outgoing
$\zeta_{-}$ and incoming $\zeta_{+}$ wave-fronts can not move faster
than the speed of light.\begin{flalign*}
 & \zeta_{-}=4\pi^{2}\Theta\left(c_{0}\Delta t'-\squigle{r}\right)e^{-i\Omega_{n}\left(\Delta t'-\frac{\squigle{r}}{c_{0}}+t_{0}\right)}\\
 & \zeta_{+}=4\pi^{2}\Theta\left(c_{0}\Delta t'+\squigle{r}\right)e^{-i\Omega_{n}\left(\Delta t'+\frac{\squigle{r}}{c_{0}}+t_{f}\right)}\end{flalign*}
We additionally go beyond the approximations which neglect revival
\cite{Quantum-Optics-Scully}, such that the components of an {}``absorbing''
pole with the contour closed over the upper half plane, are given
by\begin{flalign*}
 & \mathcal{I}_{\pm,n,z}=\left[\left(\frac{1}{\squigle{r}^{3}}-\frac{i\Omega_{n}}{\squigle{r}^{2}c_{0}}\right)\zeta_{-}-\left(\frac{1}{\squigle{r}^{3}}+\frac{i\Omega_{n}}{\squigle{r}^{2}c_{0}}\right)\zeta_{+}\right]\vec{\wp}_{z}\end{flalign*}
\begin{flalign*}
 & \mathcal{I}_{\pm,n,x}=\\
 & \left[\left(\pm i\vec{\wp}_{y}-\frac{\left(1\pm1\right)}{2}\vec{\wp}_{x}\right)\frac{\Omega_{n}^{2}}{\squigle{r}c_{0}^{2}}+\left(\vec{\wp}_{x}\mp\vec{\wp}_{y}\right)\frac{i\Omega_{n}}{\squigle{r}^{2}c_{0}}+\frac{\vec{\wp}_{x}}{\squigle{r}^{3}}\right]\frac{\zeta_{+}}{2}\\
- & \left[\left(\mp i\vec{\wp}_{y}-\frac{\left(1\pm1\right)}{2}\vec{\wp}_{x}\right)\frac{\Omega_{n}^{2}}{\squigle{r}c_{0}^{2}}-\left(\vec{\wp}_{x}\pm\vec{\wp}_{y}\right)\frac{i\Omega_{n}}{\squigle{r}^{2}c_{0}}+\frac{\vec{\wp}_{x}}{\squigle{r}^{3}}\right]\frac{\zeta_{-}}{2}\end{flalign*}
\begin{flalign*}
 & \mathcal{I}_{\pm,n,y}=\\
 & \left[\left(\mp i\vec{\wp}_{x}+\frac{\left(1\mp1\right)}{2}\vec{\wp}_{y}\right)\frac{\Omega_{n}^{2}}{\squigle{r}c_{0}^{2}}+\left(\vec{\wp}_{y}\mp\vec{\wp}_{x}\right)\frac{i\Omega_{n}}{\squigle{r}^{2}c_{0}}-\frac{\vec{\wp}_{y}}{\squigle{r}^{3}}\right]\frac{\zeta_{+}}{2}\\
- & \left[\left(\pm i\vec{\wp}_{x}+\frac{\left(1\mp1\right)}{2}\vec{\wp}_{y}\right)\frac{\Omega_{n}^{2}}{\squigle{r}c_{0}^{2}}-\left(\vec{\wp}_{y}\pm\vec{\wp}_{x}\right)\frac{i\Omega_{n}}{\squigle{r}^{2}c_{0}}-\frac{\vec{\wp}_{y}}{\squigle{r}^{3}}\right]\frac{\zeta_{-}}{2}\end{flalign*}
as governed again by the conditions that follow from the fact that
the outgoing $\zeta_{-}$ and incoming $\zeta_{+}$ wave-fronts can
not move faster than the speed of light\begin{flalign*}
 & \zeta_{-}=4\pi^{2}\Theta\left(c_{0}\Delta t'-\squigle{r}\right)e^{-i\Omega_{n}\left(\Delta t'-\frac{\squigle{r}}{c_{0}}+t_{f}\right)}\\
 & \zeta_{+}=4\pi^{2}\Theta\left(c_{0}\Delta t'+\squigle{r}\right)e^{-i\Omega_{n}\left(\Delta t'+\frac{\squigle{r}}{c_{0}}+t_{0}\right)}\end{flalign*}
The coupling between \eqref{eq:Dirac-like-photon-eqn}, \eqref{eq:interaction-negative-helicity},
and \eqref{eq:interaction-positive-helicity} was modeled computationally
through use of the following algorithm as implemented in a leap-frogging
scheme \cite{press2007numerical} between real and imaginary parts
of the wave-functions, both within and beyond the interaction regions.
\begin{itemize}
\item {\small Initialize QD excited state}{\small \par}
\item {\small Determine analytic approximation for $c_{a}$}{\small \par}
\item {\small Use the leading coefficients and roots of the analytic form
of $c_{a}$ to determine the state of the photonic wave function $\vec{\Psi}_{\gamma,+,b}^{\left(+\right)}$
\& $\vec{\Psi}_{\gamma,-,b}^{*\left(+\right)}$ at the next time step}{\small \par}
\item {\small At this new time step use the current state of the photonic
wave function to update the state of the quantum dot}{\small \par}
\item {\small Propagate $\vec{\Psi}_{\gamma,+,b}^{\left(+\right)}$ \& $\vec{\Psi}_{\gamma,-,b}^{*\left(+\right)}$}{\small \par}
\item {\small Repeat from step 2}{\small \par}
\end{itemize}
Figure \eqref{fig:Near-field-revival} shows the stable near-field
exchange of energy between the quantum dot and single photon states.
These results are derived from two computational experiments set to
have the same ratio of $\frac{\Delta x}{\Delta t}$. It is interesting
to note that the envelope functions for both of these are very similar
and both exhibit a seemingly linear decay of energy shortly after
an initial revival of the quantum dot state. This seems to suggest
that though it is worth while to retain a time resolution small enough
to observe the initial revival event, this resolution can be made
coarser almost immediately following the first revival. %
\begin{figure}
\begin{minipage}[t]{1.7in}%
\subfloat[$\left|\digamma_{z}\right|_{+}^{2}$ in XY Plane\label{fig:QD-region1}]{\centering{}\includegraphics[angle=90,width=1\columnwidth]{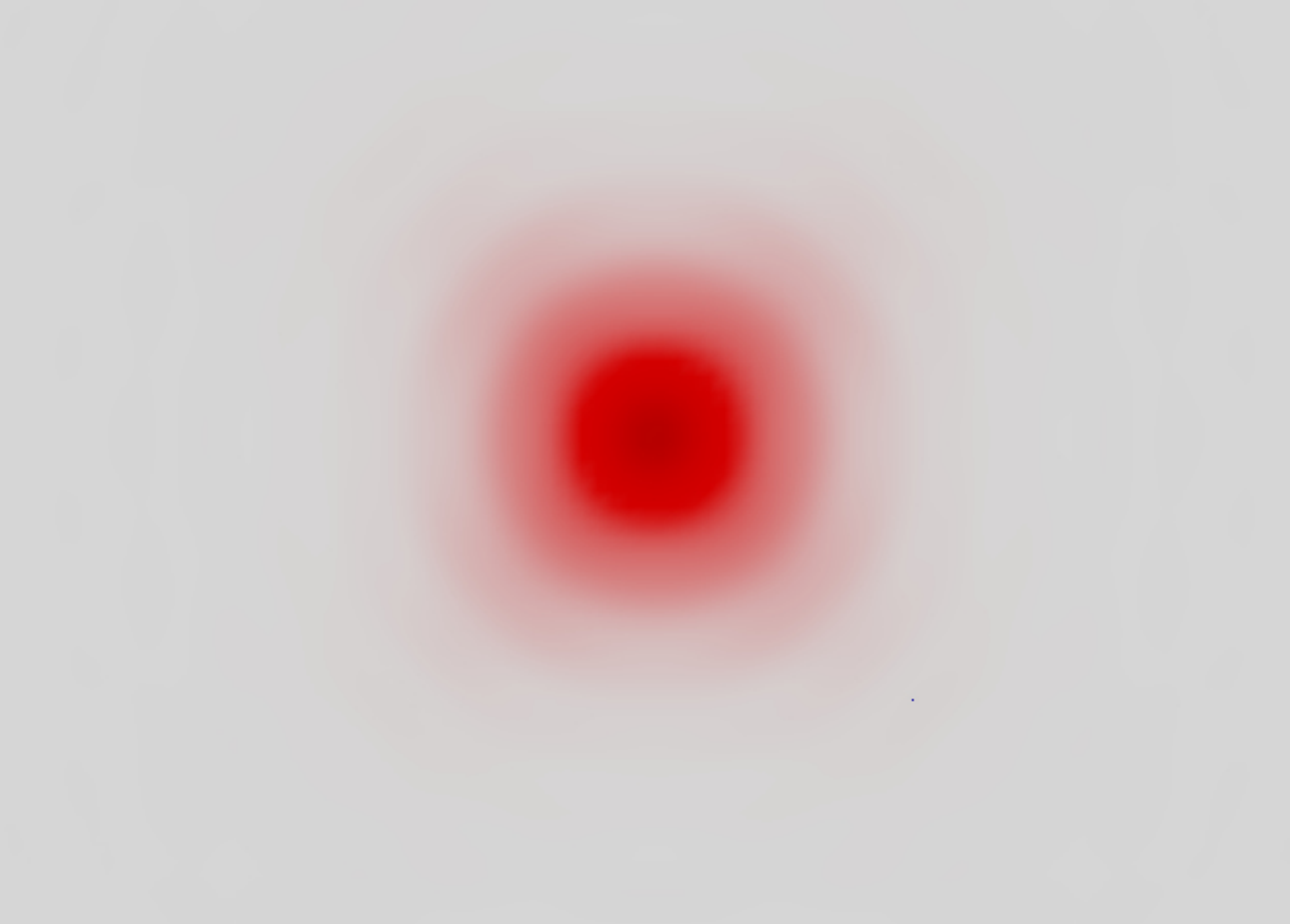}}%
\end{minipage}\hfill{}%
\begin{minipage}[t]{1.7in}%
\subfloat[$\Re\left\{ \digamma_{y}\right\} _{+}$ in XY Plane\label{fig:QD-region2}]{\centering{}\includegraphics[angle=90,width=1\columnwidth]{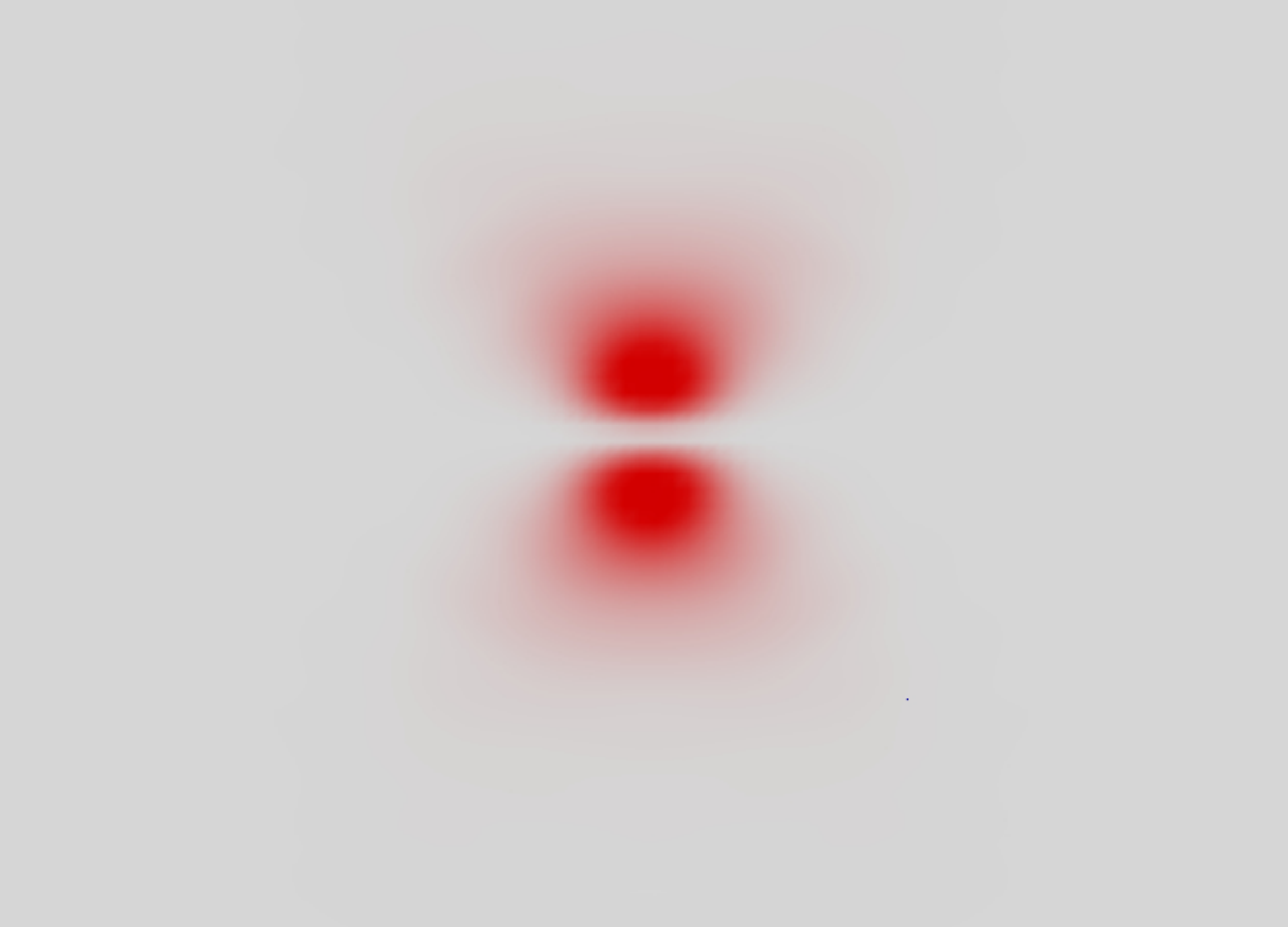}}%
\end{minipage}\hfill{}

\begin{minipage}[t]{1.7in}%
\subfloat[$\Re\left\{ \digamma_{z}\right\} _{+}$ in XY Plane \label{fig:DipolarProjection1}]{\centering{}\includegraphics[angle=90,width=1\columnwidth]{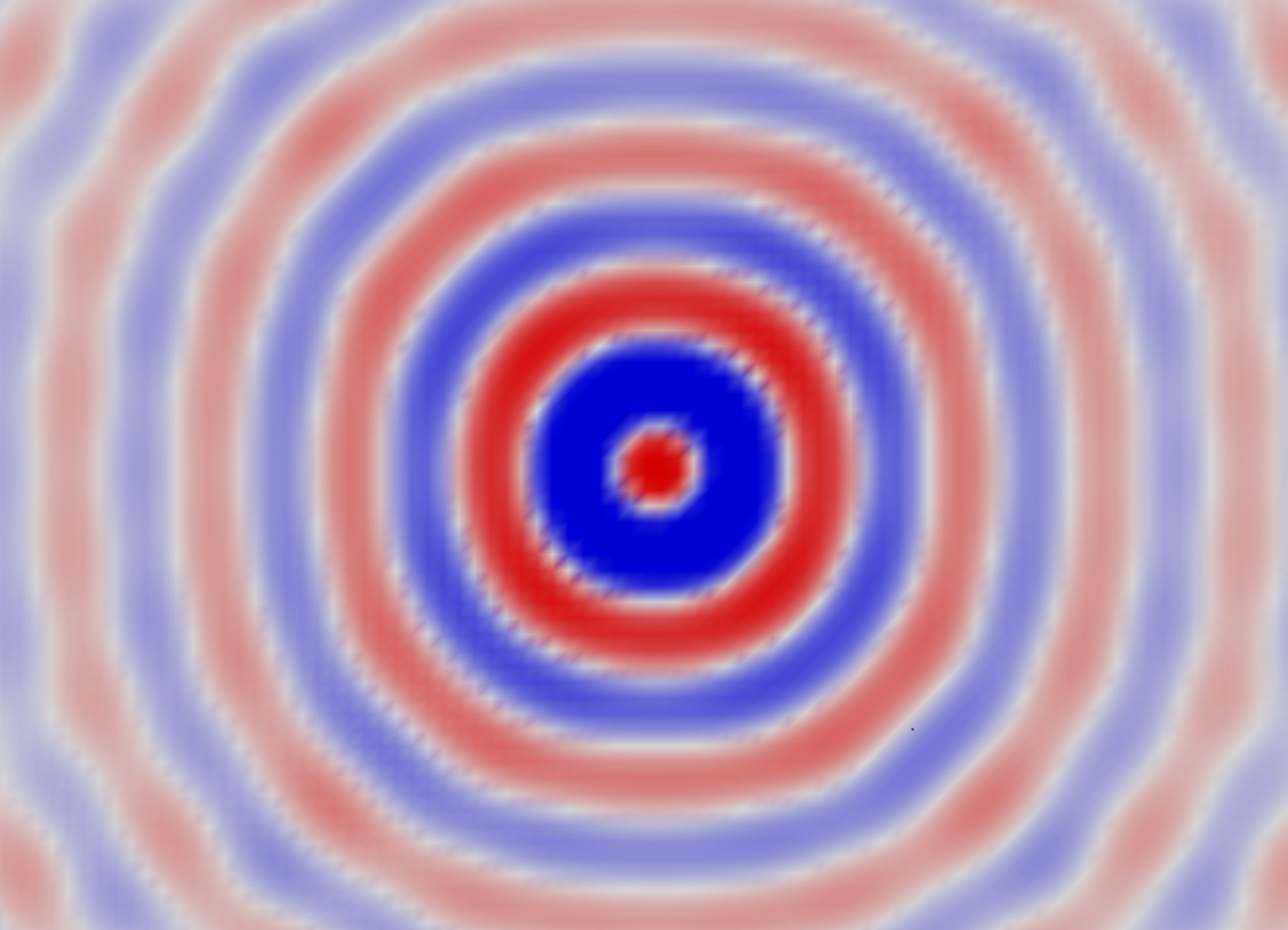}}%
\end{minipage}\hfill{}%
\begin{minipage}[t]{1.7in}%
\begin{center}
\subfloat[$\Re\left\{ \digamma_{y}\right\} _{+}$ in XY Plane]{\centering{}\includegraphics[angle=90,width=1\columnwidth]{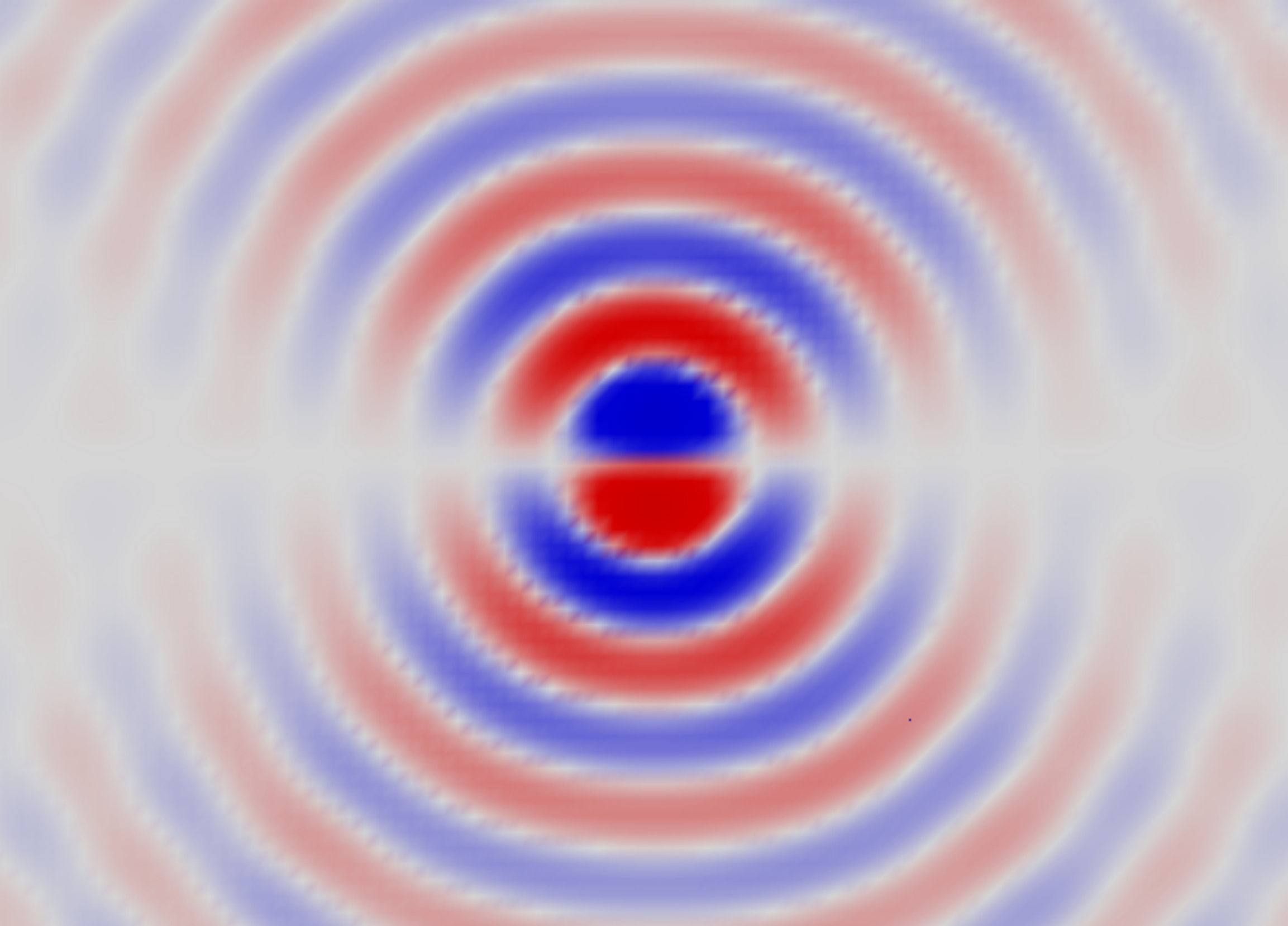}}
\par\end{center}%
\end{minipage}\hfill{}

\begin{minipage}[t]{1.7in}%
\begin{center}
\subfloat[$\Re\left\{ \digamma_{z}\right\} _{+}$ in XZ Plane\label{fig:DipolarProjection2}]{\centering{}\includegraphics[angle=90,width=1\columnwidth]{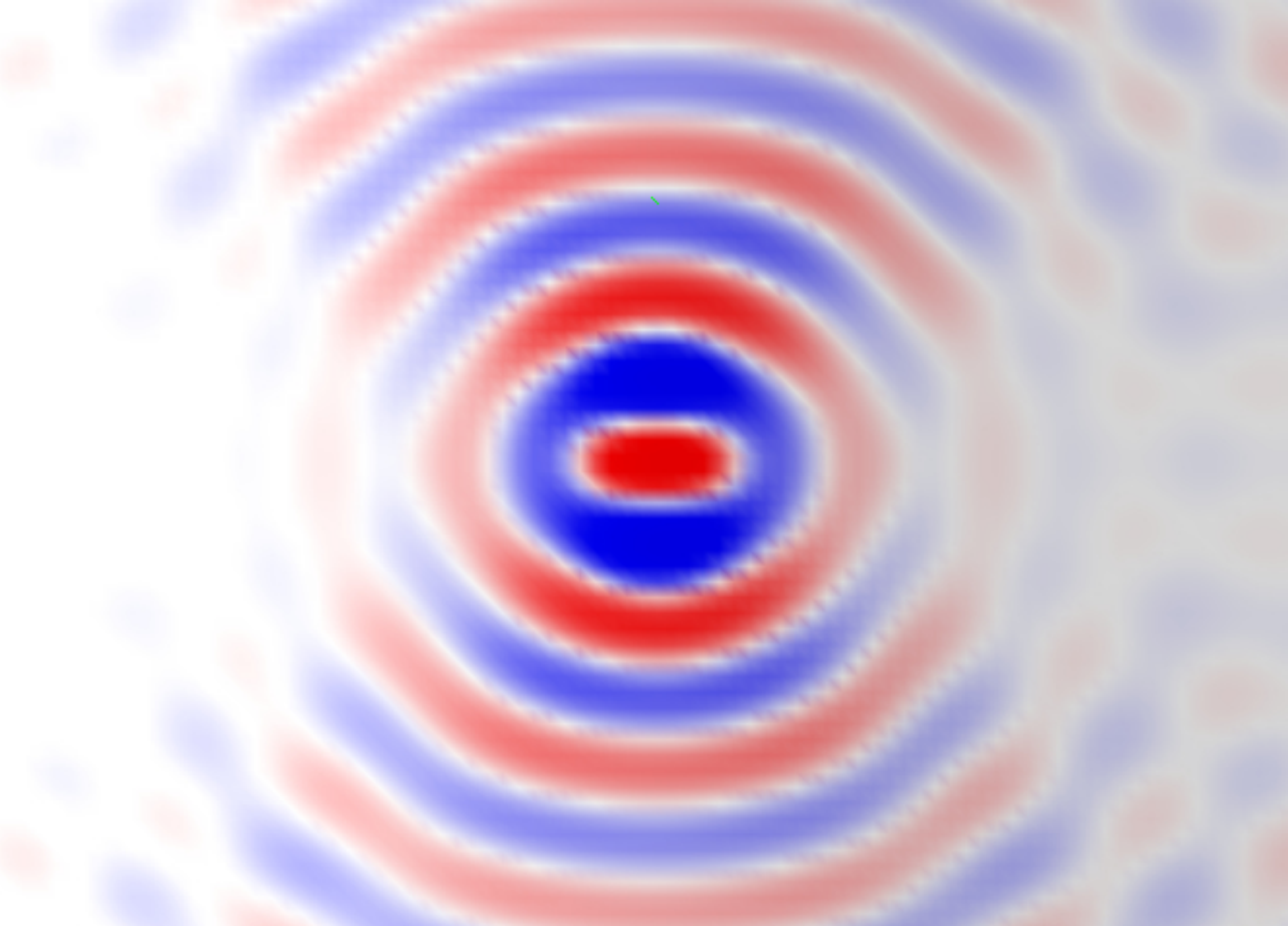}}
\par\end{center}%
\end{minipage}\hfill{}%
\begin{minipage}[t]{1.7in}%
\begin{center}
\subfloat[$\Re\left\{ \digamma_{y}\right\} _{+}$ in XZ Plane]{\centering{}\includegraphics[angle=90,width=1\columnwidth]{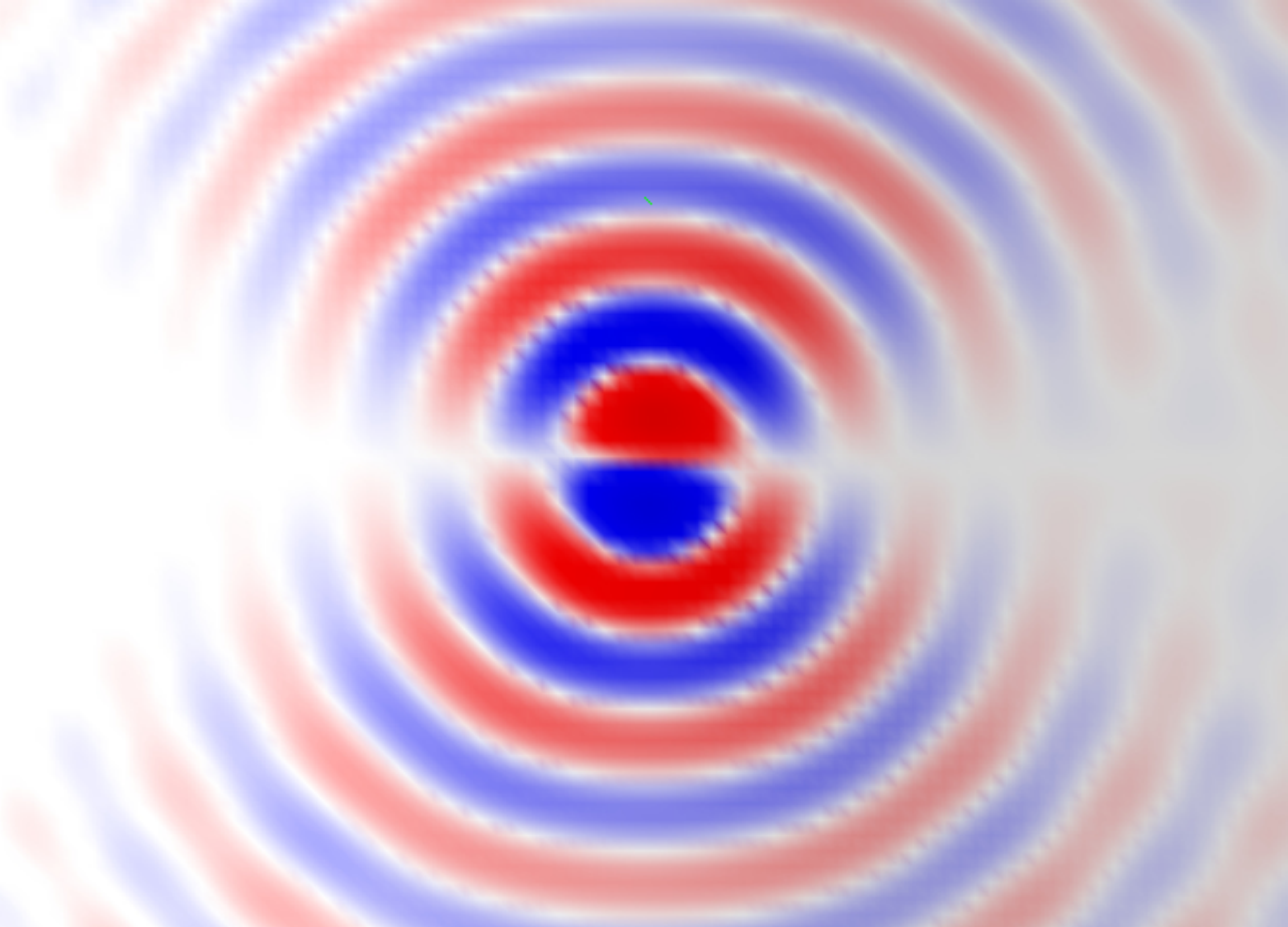}}
\par\end{center}%
\end{minipage}\hfill{}\caption{{\footnotesize Finite difference models of the spontaneous emission
of a single photon from a two-level system of quantum dot states.
Color intensity in red (positive) and blue (negative) represent the
amplitude of the photonic wave-function field strengths.\label{fig:FiniteDifferenceModels}}}

\end{figure}
 Figure \eqref{fig:FiniteDifferenceModels} presents spatial XY and
XZ plane projections of single photon state components $\digamma_{y,+}$,
$\digamma_{z+}$ and their probability amplitudes immediately following
the establishment of coherent oscillations within the polynomial envelope
functions shown in Figure \eqref{fig:Near-field-revival}. It is evident
that during this time, though evanescent modes appear to escape the
quantum dot region, calculation of the localization of the single
photon wave-function find the single photon state to be localized
to the quantum dot region in figures \eqref{fig:QD-region1} \& \eqref{fig:QD-region2}.
The dipolar structure in the projections presented by figures \eqref{fig:DipolarProjection1}
\& \eqref{fig:DipolarProjection2} was expected due to the initial
orientation solely along the z-axis of the transition dipole moment
of the two associated quantum dot levels.

We have demonstrated that it is feasible to study near field single
photon emission within dielectric structures by means of the Riemann-Silberstein
wave-function beyond the Weisskopf-Wigner approximation; both analytically
and computationally. It was further demonstrated that the locality
of a photonic state could be well described during spontaneous emission
while energy is injected and exchanged between both single photon
and quantum dot states. Test cases were directly compared for different
values of $\Delta x$ \& $\Delta t$ . From these it was determined
that the polynomial envelope functions for coherent oscillations were
in agreement within the initial revival period of the quantum dot
excited state. Furthermore, the theoretical approximations made accurately
yield analytic and intuitive insight to the periodicity of the initial
decay and revival phenomena present in the near field limit. This
work therefore makes it feasible to computationally design photonic
states to be emitted and detected by solid-state quantum dots embedded
within dielectric structures and to compare them to experimental results
by means of their corresponding density matrix and Wigner functions.

{\scriptsize \par}

\paragraph{{\scriptsize Acknowledgments}}

{\scriptsize We thank Mr. Christopher Ellis, Dr. Mikhail Erementchouck,
and Dr. Volodymyr Turkowski for helpful discussions. We acknowledge
computational support from the Institute for Simulation and Training
STOKES HPCC and financial support from NSF (Grant No. ECCS-0725514),
DARPA/MTO (Grant No. HR0011-08-1-0059), NSF (Grant No. ECCS-0901784),
AFOSR (Grant No. FA9550-09-1-0450).}{\scriptsize \par}

\paragraph*{{\scriptsize Competing Financial Interests}}

{\scriptsize The authors declare that they have no competing financial
interests.}
\end{document}